# HIGH-POWER CIRCULATOR TEST RESULTS AT 350 AND 700 MHZ

W. T. Roybal, J. Bradley III, D. Rees, P. A. Torrez,
D. K. Warner, LANL, Los Alamos, NM 87545, USA
J. DeBaca, (General Atomic) LANL, Los Alamos, NM 87545, USA

*Abstract*

The high-power RF systems for the Accelerator Production of Tritium (APT) program require high-power circulators at 350 MHz and 700 MHz to protect 1 MW Continuous Wave (CW) klystrons from reflected power. The 350 MHz circulator is based on the CERN, ESRF, and APS designs and has performed very well. The 700 MHz circulator is a new design. Prototype 700 MHz circulators have been high-power tested at Los Alamos National Laboratory (LANL). The first of these circulators has satisfied performance requirements. The circulator requirements, results from the testing, and lessons learned from this development are presented and discussed.

## 1 INTRODUCTION

The APT 350 and 700 MHz klystrons require circulators that are capable of operating indefinitely with up to full reflection at any phase. The circulators were specified to function properly under these conditions at power levels up to the maximum CW forward power out of each klystron: 1.2 MW for the 350 MHz klystron and 1 MW for the 700 MHz klystron. A 350 MHz circulator was easily produced as 352 MHz circulators at these power levels are already in use. The APT 350 MHz, 1.2 MW circulator produced by Advanced Ferrite Technology, Inc. (AFT) is shown in Figure 1.

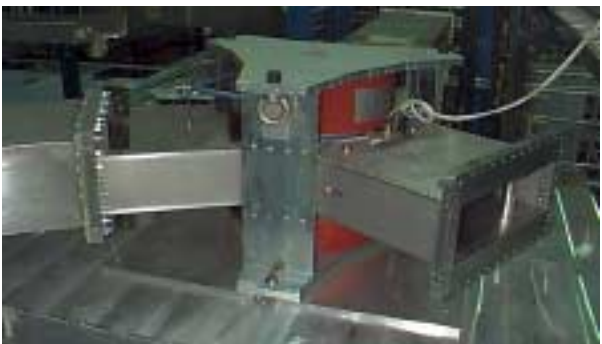

Figure 1: AFT 350 MHz APT Circulator.

Development of the 1 MW, 700 MHz circulator was more challenging. AFT was initially the sole vendor, but we chose Atlantic Microwave Corporation as a second vendor after several failures by AFT to meet the high-power operating requirements. Los Alamos National Laboratory worked closely with both AFT and Atlantic Microwave Corporation to develop and test 700 MHz prototypes for APT.

## 2 CIRCULATOR REQUIREMENTS

The APT circulator specifications for insertion loss, isolation and waveguide contention flanges for the 350 and 700 MHz circulators are shown in Table 1.

Table 1: APT Circulator Specifications

| Requirement | 350 MHz | 700 MHz |
|---|---|---|
| Insertion Loss | < 0.05 dB | < 0.05 dB |
| Isolation | > 26 dB | > 26 dB |
| Flange | WR2300 | WR1500 |

The limits on circulator loss were determined by the required control margin for the APT accelerator and the maximum saturated klystron power. These factors dictated that the sum of all waveguide and circulator losses be less than 5%. This in turn required that the budget of allowable insertion loss for the circulators be 0.05 dB at the design center frequency of the RF system.

At least 26 dB of isolation is required in order to protect the klystron from excessive reflected power while operating at megawatt CW power levels. We specified that the vendors demonstrate this ability while operating at full power with a short circuit on the output port and over the range of all possible phases of reflected power. The circulator is required to operate under all conditions of the specification without arcing. Arc detectors were required to verify that no arcing occurred and to protect the circulator in the event of arcing. Finally, the circulator is required to satisfy all requirements from startup to steady-state operating temperature without the need for operator adjustments to a Temperature Compensating Unit (TCU). A TCU is typically used to adjust the magnet coil current at the circulator, thereby maintaining the fields at the ferrite plates for maximum isolation over a range of operating temperatures.

## 3 TEST RESULTS

### 3.1 AFT 350 MHz Circulator

The AFT 350 MHz circulator was tested at power levels up to 1100 kW with a waveguide shorting plate on the output port. A shorting plate on the output port produces a standing wave in the circulator ferrite. Various lengths of waveguide were inserted between the circulator and the shorting plate to adjust the phase of the standing wave. The maximum voltage stress in the circulator occurs when this standing wave has an E-field maximum in the ferrite at the center of the circulator. We tested the voltage breakdown (arcing) characteristics by placing the E-field maximum at the center of the circulator. The maximum dissipated power in the circulator occurs when this standing wave has an H-field maximum in the ferrite at the center of the circulator. We added a quarter wave section of waveguide between the circulator and the shorting plate to obtain an H-field maximum at the center of the circulator to test the dissipated power characteristics. The data taken in the H-field maximum configuration showed the highest power dissipation in the ferrite, supporting our calculations.

As expected, the 350 MHz circulator provided by AFT met all requirements without failure. The circulator performed well and demonstrated >26 dB of isolation with less than 0.05 dB of insertion loss at all phases. Continuous operation in both the maximum E and H field configurations demonstrated reliable voltage standoff at the voltage maximum and reliable power handling capabilities at the maximum steady–state temperature. No arcs were detected over the full range from turn on to steady-state operation. Once the TCU was properly adjusted to the optimum setting, the entire set of tests were performed without requiring further adjustment.

### 3.2 AFT 700 MHz Circulators

AFT has delivered three 1.0 MW, 700 MHz circulators for use on the Low Energy Demonstration Accelerator (LEDA). All three circulators have been tested on a test stand and each one has had difficulty meeting the final high-power acceptance test criteria. The problems encountered included arcing, poor isolation, and poor TCU performance. A total of 5 high-power testing sessions were required for the first circulator to meet all specifications. As a result of this performance, we selected Atlantic Microwave as an alternate vendor.

AFT initially approached this development by scaling down the well-demonstrated 350 MHz WR2300 circulator to be packaged in WR1500 waveguide and operate equivalently at 700 MHz. The prototype met all low power requirements on the bench but had problems under high power. A picture of one of the three AFT 700 MHz circulators delivered to LANL is shown in Figure 2.

After demonstrating that it could pass all low power tests at AFT, the AFT 700 MHz prototype circulator was shipped to Los Alamos for high-power testing. We obtained a sliding short device from Mega Industries for the full-power, full-reflection circulator tests. This sliding short installed on the output port of the circulator allowed us to vary the RF phase of the short continuously over a range more than 360 degrees.

The first test resulted in continuous arcing at forward power levels greater than 275 kW at E-field maximum. During the testing AFT personnel determined that there was an incorrect cooling water distribution between the ferrite plates and the reduced-height waveguide. Attempts were made at Los Alamos to restrict the coolant flow to the waveguide and matching posts in order to increase coolant flow to the ferrite plates. Results from these efforts were positive but the circulator continued to arc at power levels of 500 kW.

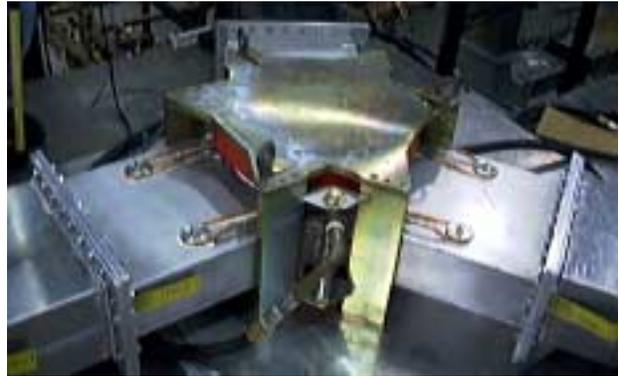

Figure 2: AFT 700 MHz APT Circulator Installed on the Test Stand.

The circulator was rebuilt for the second set of high-power tests. The second test showed improvement over the first test; however, arcing again limited the circulator to just over 500 kW at E-field maximum. Internal inspection revealed arcing on the threads of a bolt used to attach the ferrite plates to the water connection and to the waveguide structure. A major redesign and rebuild followed these findings. The circulator was fitted with a new design for supporting the plates in place and connecting the water lines to the ferrite plates.

For the third test, it seemed that the arcing issue was resolved. The circulator was tested to full power without arcing at almost all phases. Only at the H-field maximum the isolation did not meet specification. Over the range of sliding short positions ±15 degrees about the H-field maximum postion, the transmitter's klystron-reflected-power interlock tripped off the high voltage and did not allow further testing. Manual adjustments could be made to the TCU to prevent the tripping off of the system, but the specificaton required operation without manual adjustment of the TCU. It was also determined that the speed of the adjustment of the sliding short contributed to the running away of the reflected power to the klystron. The TCU needed to respond more quickly to sudden changes in phase.

The focus was then turned to upgrading the capabilities of the TCU. Improvements such as the addition of a fast-feed-forward loop were added to the electronics to obtain better performance. The fourth test demonstrated the 26 dB of isolation required, but the circulator arced under full power at the H-field max. The arc spot was located on the top plate and on the waveguide wall. The plate was replaced and the waveguide wall arc damage was cleaned for the fifth high-power test.

The fifth and final AFT circulator high-power acceptance test took place without incident. The circulator passed all requirements for insertion loss and isolation at high power when presented with a variable sliding short. The sliding short was varied in both directions at all possible speeds. Reflected power at the klystron did not exceed 2 kW and no arcs were detected. A plot of the power dissipated in the circulator versus the phase of the waveguide short during the final AFT test is given in Figure 3. Full power from the 1.0 MW klystron combined with waveguide losses, circulator losses, sliding short losses, and accuracy of power meters and calibrations yielded just over 900 kW of forward power delivered through the circulator, reflected off the sliding short, and dissipated at the high-power water load.

The two remaining AFT prototype circulators are scheduled for rebuilding for better high-voltage standoff and the TCUs will be upgraded to be equivalent to the TCU that was used with the circulator that passed final acceptance.

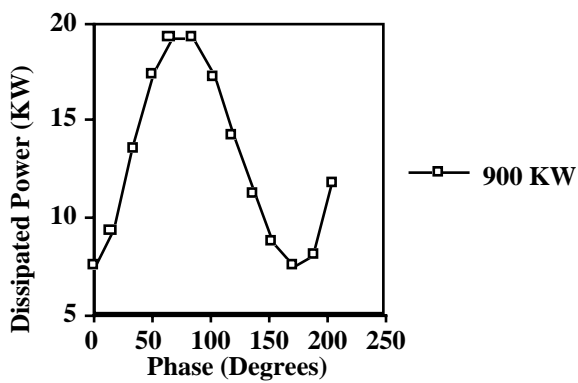

Figure 3: Plot of Dissipated Power versus Phase for the AFT 700 MHz Circulator Tested at LANL.

### 3.3 Atlantic Microwave 700 MHz Circulator

After the second failed test of the AFT 700 MHz circulator, we had Atlantic Microwave begin to develop a circulator for APT to serve as a backup source in the event that AFT could not resolve their 700 MHz circulator problems. The first part of the 3-part development plan was to design the circulator. The second part was to develop and build a circulator with a single plate of ferrite material that would be rated for 100 kW CW at 700 MHz. The third part would be to deliver a full megawatt CW circulator prototype.

Atlantic Microwave had completed the design and delivered the single plate 100 kW circulator to LANL by the time of the AFT circulator failed its fourth test. Figure 4 shows the Atlantic Microwave 100 kW prototype circulator installed on the test stand.

The Atlantic Microwave circulator performed well at its 100 kW power rating and passed all of our testing requirements at this power level. The circulator was then tested at power levels above its 100 kW rating. First we adjusted the sliding short to produce an E-field maximum in the center of the circulator and raised the RF power in 50 kW steps. The circulator was allowed to come to steady-state temperature at each 50 kW step. The circulator began to arc when the power reached 600 kW. The circulator was then tested with the sliding short adjusted to produce an H-field maximum in the center of the circulator and with the RF power level increased in 50 kW steps as before. Under these conditions, the circulator power reached only 300 kW before the circulator isolation degraded to the point that further increases in power were not possible. Improvements to the isolation can be made by using a TCU to compensate for temperature effects in the ferrite. The 100 kW circulator prototype did not include a TCU, but the design of the full-scale 1.0 MW circulator will include a TCU.

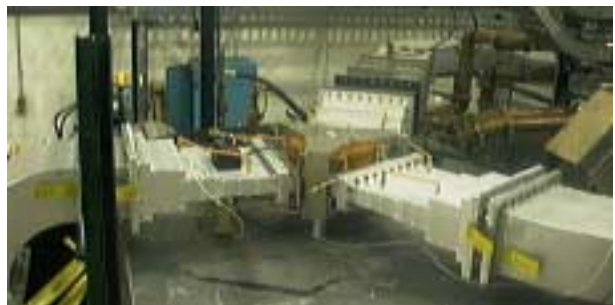

Figure 4: Atlantic Microwave 700 MHz 100 kW Circulator Installed on the Test Stand.

## 4 CONCLUSIONS

The development of a 700 MHz circulator for APT proved to be far more challenging than the development of the 350 MHz circulator. Simply scaling down the well demonstrated 350 MHz WR2300 circulator resulted in a prototype that met all low power requirements on the bench but did not meet our specifications at high power.

Rigorous testing procedures revealed inherent arcing and isolation problems in the initial prototype AFT circulator and led to substantial changes to the internal design. The lessons learned in correcting the arcing problems in high field regions within the circulator structure led to the redesign of the circulator interior for improved voltage stand off. Lessons learned in providing adequate temperature compensation at higher response rates led to the development of a faster, smarter Temperature Compensating Unit. The problems encountered in the testing phase of the AFT circulator also led us to develop an alternative supplier of high power circulators.

The high power testing results give us confidence that we can obtain a fully functional and reliable circulator for every high-power RF system required for APT. A second supplier also provides a safety margin in meeting the challenging production and delivery deadlines required if the full APT accelerator becomes fully funded and hundreds of these circulators must be produced.